\documentclass[prd,twocolumn,amsmath,showpacs,amssymb,superscriptaddress,floatfix,nofootinbib,9pt]{revtex4}
\usepackage{indentfirst}
\usepackage{tikz}
\usepackage{setspace}
\usepackage{amsfonts,color,xcolor,graphicx,amsfonts,multirow,amsmath}
\usepackage{amssymb}
\usepackage[section]{placeins}
\usepackage{booktabs}
\definecolor{bluee}{rgb}{0,0,1}
\usepackage[colorlinks=true, citecolor=bluee, linkcolor=bluee, urlcolor=bluee]{hyperref}
\definecolor{darkerblue}{rgb}{0.1, 0.1, 0.7}
\definecolor{darkergreen}{rgb}{0.1, 0.5, 0.1}
\definecolor{lightred}{rgb}{1, 0, 0}
\usepackage{ulem}

\allowdisplaybreaks
\usepackage{slashed}

\begin{document}
\title{Characterize the properties of $D_s^+$-meson decay constant and leptonic decays by using QCD sum rules within background field theory framework}
\author{Jian-Qi Chen}
\address{Department of Physics, Guizhou Minzu University, Guiyang 550025, P.R.China}
\author{Ya-Xiong Wang}
\address{School of Physical Science and Technology, Southwest University, Chongqing 400715,PR China}
\author{Hai-Bing Fu}
\email{fuhb@gzmu.edu.cn}
\address{Department of Physics, Guizhou Minzu University, Guiyang 550025, P.R.China}

\date{\today}

\begin{abstract}
The leptonic decays of the $D_s^+$-meson have received considerable attention in recent years. In this work, we perform a precise calculation of the decay constant $f_{D_s^+}$ using the QCD sum rules method within the background field theory framework. In our calculation, we fully include the quark propagator contributions up to dimension-six condensates. By adopting two different constraint schemes, we obtain $f_{D_s^+}^{\text{(I)}} = 253.0_{-3.1}^{+3.3}\ \text{MeV}$ and $f_{D_s^+}^{\text{(II)}} = 251.8_{-1.3}^{+1.4}\ \text{MeV}$, respectively, both of which are in good agreement with existing theoretical and experimental results. The conventional scheme follows the standard Borel window criteria, while the derivative scheme reduces the dependence of the decay constant on the Borel parameter through an auxiliary function. Based on these decay constants and incorporating the NLO electroweak radiative corrections, we further calculate the branching fractions for the three leptonic decay channels for both schemes. Combined with the latest branching fraction $\mathcal{B}(D_s^+ \to \mu^+ \nu_\mu)$ from the PDG, we extract the CKM matrix elements $|V_{cs}|^{\text{(I)}} = 0.967 \pm 0.012$ and $|V_{cs}|^{\text{(II)}} = 0.970 \pm 0.005$ from the two schemes, respectively.
\end{abstract}

\maketitle
\textit{Introduction.--}The leptonic decay of the $D_s^+$-meson provides an ideal platform for the precise extraction of the Cabibbo--Kobayashi--Maskawa (CKM) matrix element $|V_{cs}|$. This decay proceeds via the annihilation of the $c$-quark and the $\bar{s}$-quark into a virtual $W^+$ boson, which subsequently yields the $\ell^+\nu_\ell$ final state. Among the three lepton flavors, the $e^+\nu_e$ mode is severely suppressed due to helicity suppression, making it difficult to measure experimentally and only upper limits are currently available. In contrast, the branching fractions of the $\mu^+\nu_\mu$ and $\tau^+\nu_\tau$ modes have been measured with good precision. Together with the $D_s^+$ lifetime and the decay constant $f_{D_s^+}$, these measurements provide a clean and reliable way to extract $|V_{cs}|$. Therefore, a theoretical prediction of the $D_s^+$-meson decay constant can provide useful references for experimental measurements and test the Standard Model.

Over the past two decades, significant experimental studies have been devoted to measuring $f_{D_s^+}$ with increasing precision. In 2008, the Belle Collaboration measured $f_{D_s^+} = 275 \pm 16_{\rm stat} \pm 12_{\rm syst}\ \text{MeV}$ using a $548\ \text{fb}^{-1}$ data sample collected at the KEKB $e^+e^-$ collider~\cite{Belle2008}. In 2009, the CLEO-c Collaboration reported a determination of $f_{D_s^+} = 252.5 \pm 11.1 \pm 5.2\ \text{MeV}$ from a sample of 26,334 tagged $D_s^+$ decays via the $D_s^+ \to \tau^+\nu_\tau$ channel~\cite{CLEO0901a}. Besides, they performed a second independent determination, obtaining $f_{D_s^+} = 263.3 \pm 8.2 \pm 3.9\ \text{MeV}$ from a combined analysis of the $D_s^+ \to \mu^+\nu_\mu$ and $D_s^+ \to \tau^+\nu_\tau$ branching fraction measurements~\cite{CLEO0901b}. In 2021, the BESIII Collaboration, using a larger data sample and improved analysis methods together with the $|V_{cs}|$ value from the CKMfitter group, obtained $f_{D_s^+} = 251.1 \pm 2.4 \pm 3.0\ \text{MeV}$~\cite{BESIII2106}. These results indicate that the discrepancy among different experimental determinations of $f_{D_s^+}$ remains unresolved, underscoring the need for further improvements in both experimental techniques and theoretical calculations.

 On the theoretical side, substantial progress has also been made in determining $f_{D_s^+}$ using various non-perturbative methods. Lattice QCD has been extensively applied to the calculation of heavy-light meson decay constants over the past two decades \cite{Chiu2005,Aubin2005,Chen2008,Blossier2014,Blossier2018,Du2024,HPQCD0706,LQCD1712,FLAG2411}. In 2007,  the HPQCD and UKQCD Collaborations calculated the $D_s^+$ decay constant using the highly improved quark discretisation on MILC gluon field, obtaining $f_{D_s^+} = 241 \pm 3\ \text{MeV}$~\cite{HPQCD0706}. By 2017, the Fermilab Lattice and MILC Collaborations substantially improved the theoretical precision using finer lattices and more reliable continuum extrapolations, obtaining $f_{D_s^+} = 249.9(4)\ \text{MeV}$ \cite{LQCD1712}. In 2020, the $\chi$QCD Collaboration determined $f_{D_s^+} = 249(7)$~MeV using $2+1$-flavor lattice QCD at the physical pion mass, extracting the value from pseudoscalar density matrix elements~\cite{Chen2008}. Different from lattice QCD, the QCD sum rules approach provides an independent theoretical framework based on the operator product expansion and Borel transformation \cite{Gelhausen2013,Lucha2010Ds,Wang2015a,Mehraban2018Ds}. In 2010, a systematic study of the Borel parameter dependence of the effective continuum thresholds was performed, yielding $f_{D_s^+} = 245.3 \pm 15.7 \pm 4.5\ \text{MeV}$ \cite{Lucha2010Ds}. In 2015, the result $f_{D_s^+} = 240 \pm 10\ \text{MeV}$ was obtained with only the $\mathcal{O}(\alpha_s)$ corrections, while including the $\mathcal{O}(\alpha_s^2)$ corrections with the pole mass gave $f_{D_s^+} = 259 \pm 10\ \text{MeV}$, demonstrating the importance of higher-order perturbative contributions~\cite{Wang2015a}. In 2018, a simple interpolating current was employed within the QCD sum rules framework, taking the massless limit for light quarks, and the value $f_{D_s^+} = 245.70 \pm 7.46\ \text{MeV}$ was obtained~\cite{Mehraban2018Ds}. Despite these efforts, discrepancies persist among different theoretical approaches and a fully consistent value of $f_{D_s^+}$ has yet to be established, motivating further refinements in theoretical calculations of $f_{D_s^+}$.

The CKM matrix element $|V_{cs}|$ has been determined by several experimental and theoretical groups using a variety of decay channels. In 2009, the CLEO-c Collaboration obtained $|V_{cs}| = 1.019(10)(7)(106)$ from an analysis of $D \to K e^+\nu_e$ semileptonic decays combined with lattice QCD form factors \cite{CLEO2009D}. Over the past decade, the BESIII Collaboration has made significant progress in measuring $|V_{cs}|$ \cite{BESIII2015,BESIII1901,BESIII2102,BESIII2105,BESIII2606,BESIII1608,BESIII2407}. In 2015, the BESIII Collaboration measured the absolute branching fraction $\mathcal{B}(D^+ \to \bar{K}^0 e^+\nu_e)$, determined $f_+^K(0)|V_{cs}|$ from the differential decay rate and extracted $|V_{cs}| = 0.975 \pm 0.008 \pm 0.015 \pm 0.025$ using unquenched lattice QCD calculations for $f_+^K(0)$ as input \cite{BESIII2015}. In 2019, based on an analysis of a $3.19\ \text{fb}^{-1}$ data sample, they measured $|V_{cs}| = 1.031 \pm 0.012 \pm 0.009 \pm 0.079$ from the $D_s^+ \to \eta^{(\prime)} e^+\nu_e$ decay channels \cite{BESIII1901}. Compared with semileptonic decays, which require the $q^2$-dependent form factors $f_+(q^2)$ as inputs, leptonic decays depend only on the decay constant $f_{D_s^+}$, providing a cleaner extraction of $|V_{cs}|$. In 2021, the BESIII Collaboration determined $|V_{cs}| = 0.973 \pm 0.012 \pm 0.015 \pm 0.004$ from the $D_s^+ \to \mu^+\nu_\mu$ decay channel \cite{BESIII2102} and $|V_{cs}| = 0.980 \pm 0.023 \pm 0.019$ from the $D_s^+ \to \tau^+\nu_\tau$ decay channel \cite{BESIII2105}. In 2026, they combined four $\tau^+$ decay modes to measure $\mathcal{B}(D_s^+ \to \tau^+\nu_\tau) = (5.37 \pm 0.08 \pm 0.06)\%$ and extracted $|V_{cs}| = 0.993 \pm 0.008 \pm 0.006 \pm 0.003 \pm 0.003$~\cite{BESIII2606}.
On the theory side, using the continuum Schwinger function methods, Ref.~\cite{Yao2021} obtained parameter-free predictions for the $D_s \to K$ semileptonic form factors and branching fractions in 2021, yielding $|V_{cs}| = 0.974(10)$. In 2025, a complete one-loop calculation of the electroweak and QED corrections to the $D_s^+ \to \mu^+\nu_\mu$ and $D_s^+ \to \tau^+\nu_\tau$ decays was performed, and $|V_{cs}| = 0.991 \pm 0.007$ was obtained from the latest experimental data~\cite{Kitahara2025}. Based on the above discussions, it can be seen that the extracted $|V_{cs}|$ values from different channels are generally consistent within uncertainties, while their precision remains predominantly limited by the theoretical uncertainty in the decay constant. To effectively address the non-perturbative problems involved in the determination of the decay constant, we use the QCD sum rules approach to calculate the $D_s^+$ decay constant.

The QCD sum rules approach provides a well-established non-perturbative framework for studying hadron properties, relating observable quantities to fundamental QCD parameters through the operator product expansion and Borel transformation \cite{SVZ1979a,SVZ1979b,SVZ1979c}. In the background field theory framework, we fully retain the quark propagator contributions up to dimension-six condensates and systematically incorporate the $s$-quark mass effect, leading to a more complete and reliable determination of $f_{D_s^+}$. On this basis, we propose two different constraint schemes for comparison. The first scheme follows the conventional criteria based on pole dominance, OPE convergence and the a stable Borel window. The second scheme, based on the derivative of the auxiliary function with respect to $1/M^2$, provides an alternative constraint that does not require prior assumptions on the Borel window boundaries, thereby reducing the subjective uncertainty in the window selection. By applying these two constraint schemes, we significantly reduce the sensitivity of $f_{D_s^+}$ to the Borel parameter and obtain a more precise Borel window. These improvements enable a more accurate determination of the $D_s^+$-meson decay constant $f_{D_s^+}$, which will be used to extract $|V_{cs}|$ from the experimentally measured branching fractions in the subsequent analysis, thereby providing a valuable reference for probing new physics beyond the Standard Model.
%

\textit{Theoretical Framework.--}
In the framework of QCD sum rules, the decay constant $f_{D_s^+}$ is obtained from the definition involving the moments of the twist-2 distribution amplitude of the $D_s^+$ meson. The 0th moment satisfies the normalization condition~\cite{Zhang2018BD}, and this definition simplifies to the standard matrix element form:
\begin{equation}
\langle 0|\bar{c}(0)\gamma_\mu\gamma_5 s(0)|D_s^+(q)\rangle = i q_\mu f_{D_s^+}.
\end{equation}
Then we can construct the two-point correlation function. We establish the connection between fundamental QCD parameters and hadronic observables and subsequently derive the QCD sum rule expression for $f_{D_s^+}$:
\begin{align}
\label{correlator1}
\Pi_{D_s^+}(z,q) &= i \int d^4x\, e^{iq\cdot x} \langle 0| T\{J_n(x)J_0^\dagger(0)\}|0\rangle
\nonumber\\
& = (z\cdot q)^{2} I_{D_s^+}(q^2),
\end{align}
 To effectively suppress higher-twist contributions and excited-state contaminations, while simplifying the OPE calculations within the background field theory, we adopt the interpolating current of the following form:
\begin{align}
J(x) = \bar{c}(x)\slashed{z}\gamma_5 s(x),
J^\dagger(0) = \bar{s}(0)\slashed{z}\gamma_5 c(0).
\end{align}
In the deep Euclidean region, the non-perturbative long-distance effects need to be treated properly. In the background field theory, the non-perturbative vacuum effects are incorporated into the quark propagator by expanding it in powers of the background gluon field, which systematically generates the contributions from vacuum condensates of increasing dimensions. The OPE of the correlation function~\eqref{correlator1} then gives:
\begin{align}
\Pi_{D_s^+}(z,q) & = i \int d^4x\, e^{iq\cdot x}
\nonumber\\
&\times \big\{ -\mathrm{Tr}\langle 0| S_F^c(0,x)\slashed{z}\gamma_5 S_F^q(x,0)\slashed{z}\gamma_5 |0\rangle \nonumber\\
&+\mathrm{Tr}\langle 0| \bar{c}(x)c(0)\slashed{z}\gamma_5 S_F^q(x,0)\slashed{z}\gamma_5 |0\rangle \nonumber\\
&+\mathrm{Tr}\langle 0| S_F^c(0,x)\slashed{z}\gamma_5 \bar{q}(0)q(x)\slashed{z}\gamma_5 |0\rangle
\nonumber\\
& + \cdots \big\},
\label{correlator2}
\end{align}
where $\mathrm{Tr}$ denotes the trace of \(\gamma\)-matrices and color matrix. $S_F^c(0,x)$ and $S_F^q(x,0)$ are the quark propagators in the background field, whose explicit expressions for all terms can be found in our previous work~\cite{Huang1989BFT,Zhong2014PionDA,Hu2022Eta}. Here we present their expansion up to dimension-six operators:
\begin{align}
\label{quarkpropagator}
S_F(x,0) & = S_F^0(x,0) + S_F^2(x,0) + S_F^3(x,0)
\nonumber\\
&+ \sum_{i=1}^2 S_F^{4(i)}(x,0)
+ \sum_{i=1}^3 S_F^{5(i)}(x,0)
\nonumber\\
&+ \sum_{i=1}^5 S_F^{6(i)}(x,0).
\end{align}

Moreover, we notice that the infrared divergence arises from the non-convergence of the integral in the region where the light-quark momentum tends to zero in the limit of vanishing light-quark mass. To regularize the infrared divergences appearing in the leading terms, we adopt dimensional regularization with $D = 4 - 2\epsilon$ ($\epsilon \to 0$) in combination with the Feynman parametrization formula and finally obtain:
\begin{align}
I(m,a,b,c) &= \sum_{k=0}^{m} \frac{(-1)^k m!}{k!(m-k)!}
\int_0^1 dx\, x^{m-k-a k\epsilon}\bar x^{k+b}
\nonumber\\
&\times \left(1 - \frac{q^2}{-q^2 + m_c^2}\right)^{-c-\epsilon}.
\end{align}
This integral can be simplified using the hypergeometric function:
\begin{align}
F(\alpha,\beta,\gamma,Z) &= \frac{\Gamma(\gamma)}{\Gamma(\beta)\Gamma(\gamma-\beta)}
\int_0^1 dx\, x^{\beta-1}\bar x^{\gamma-\beta-1}
\nonumber\\
& \times (1-Zx)^{-\alpha}
\nonumber\\
& = \sum_{l=0}^{\infty} \frac{(\alpha)_l(\beta)_l}{l!\,(\gamma)_l} Z^l,
\end{align}
where $|Z|<1$ and $(\lambda)_l = \Gamma(\lambda+l)/\Gamma(\lambda)$. Further, we obtain
\begin{align}
I(m,a,b,c) &= \sum_{k=0}^{m} \frac{(-1)^k m!}{k!(m-k)!}
\frac{\Gamma(k+b+1)}{\Gamma(c+\epsilon)}
\nonumber\\
&\times \sum_{l=0}^{\infty}
\frac{\Gamma(l+c+\epsilon)\Gamma(l+m-k-a+1-\epsilon)}
{\Gamma(l+m-a+b+2-\epsilon)} \notag \\
& \times \left(\frac{-q^2}{-q^2 + m_c^2}\right)^l.
\end{align}
In the OPE calculation, the infrared divergence appears in $\Gamma(l+m-k-a+1-\epsilon)$ at the lowest several $l$-terms. We adopt the $\overline{\mathrm{MS}}$ scheme to deal with these divergences and renormalize them into the decay constant. Furthermore, we note a significant mass difference between the $D_s^+$-meson and the $D^+$-meson. Since the $s$-quark is considerably heavier than the $d$-quark, the massless approximation used for $D^+$-meson would introduce a substantial deviation in the theoretical prediction of $f_{D_s^+}$ if retained. Therefore, the $s$-quark mass must be included in our calculation. On the other hand, the OPE of the correlation function~\eqref{correlator1} and its hadronic expansion in the deep Euclidean region are matched via the dispersion relation, followed by the Borel transformation. Finally, we obtain the QCD sum rule expression for the $D_s^+$-meson decay constant:
\begin{widetext}
\begin{align}
&f_{D_s^+}^2(s,M^2) = I_{\rm pert}(s,M^2) + I_{\langle \bar qq\rangle}(M^2) + I_{\langle G^2\rangle }(M^2) + I_{\langle \bar qGq\rangle }(M^2) + I_{\langle \bar qq\rangle^2}(M^2) + I_{\langle G^3\rangle }(M^2),
\label{Eq:SR_total}
\\
&I_{\rm pert}(s,M^2) = \int_{l_{\rm min}}^{s_0^{D_s^+}} \frac{ds}{16 \pi ^2 s^3} \exp \bigg( \frac{m_{D_s^+}^2 - s}{M^2} \bigg) \{ (m_s^2 - m_c^2 + {\cal C}{\cal D})[3s^2 - (m_s^2 - m_c^2 - {\cal C}{\cal D})^2] - ({\cal C}{\cal D} \leftrightarrow  - {\cal C}{\cal D})\},
\label{Eq:SR_pert}
\\
&I_{\langle \bar qq\rangle}(M^2) = -\frac{m_q\langle \bar qq\rangle}{M^2}\exp\bigg( \frac{m_{D_s^+}^2 - m_c^2}{M^2} \bigg),
\label{Eq:SR_qq}
\\
&I_{\langle G^2\rangle}(M^2) = \frac{\langle \alpha_s G^2\rangle}{12\pi M^2}\exp\bigg( \frac{m_{D_s^+}^2}{M^2} \bigg)\bigg[ \mathcal{H}(0,0,0) - \frac{m_c^2}{M^2}\mathcal{H}(0,1,-2) \bigg],
\label{Eq:SR_GG}
\\
&I_{\langle \bar qGq\rangle}(M^2) = -\frac{m_q\langle g_s\bar q\sigma TGq\rangle}{9M^4}\exp\bigg(\frac{m_{D_s^+}^2 - m_c^2}{M^2}\bigg)\bigg(\frac{1}{2} + \frac{2m_c^2}{M^2}\bigg),
\label{Eq:SR_qGq}
\\
&I_{\langle \bar qq\rangle^2}(M^2) = \frac{2\langle g_s\bar qq\rangle^2}{81M^4}\exp\bigg(\frac{m_{D_s^+}^2 - m_c^2}{M^2}\bigg),
\label{Eq:SR_qq2}
\\
&I_{\langle G^3\rangle}(M^2) = \frac{\langle g_s^3fG^3\rangle}{\pi^2M^4} \, \exp\bigg(\frac{m_{D_s^+}^2}{M^2}\bigg) \, \bigg\{ \exp\bigg(  \, -  \, \frac{m_c^2}{M^2}\bigg)  \,  \bigg\{- \frac{17}{96}{\cal F}_1(0,5,3,2,\infty)  \, - \,  \frac{1}{96}{\cal F}_2(0,4,3,1,\infty) + \frac{1}{144}
\nonumber
\\
&\qquad \qquad~ \times {\cal F}_2(0,3,3,1,\infty)~ +~\frac{1}{288} \, \bigg(50 + 51\frac{m_c^2}{M^2}\bigg) \, \bigg[\ln\frac{M^2}{\mu^2}  \, +  \, \psi(3)\bigg] \,  + \,  \frac{17 m_c^2}{288 M^2} \, \bigg\}  \, + \,  \bigg\{  \, \frac{1}{288} \, \bigg[ \, 4 \, {\cal H}(0,0,0)
\nonumber
\\
&\qquad \qquad~ - 3  {\cal H}(0,0,-1)  \, - \,  51{\cal H}(0,1,-2) \, \bigg]
 \,\, +  \, \frac{m_c^2}{288M^2} \, \bigg[-2{\cal H}(0,0,-2)  \, + \, 4{\cal H}(0,0,-1)  \, - \,  2 \, {\cal H}(-1,1,-2)
\nonumber
\\
&\qquad \qquad~
- 3{\cal H}(0,1,-3)\bigg] + \frac{1}{240}\frac{m_c^4}{M^4}{\cal H}(0,1,-4)\bigg\}\bigg\},
\label{Eq:SR_GGG}
\end{align}
\end{widetext}
where ${\cal C} = s - (m_c - m_s)^2$, ${\cal D} = \sqrt{1 - 4m_c m_s/{\cal C}}$ and $s_0^{D_s^+}$ denotes the continuum threshold parameter. The auxiliary functions $\mathcal{F}_1$, $\mathcal{F}_2$ and $\mathcal{H}$ appearing in Eqs.~\eqref{Eq:SR_GG} and \eqref{Eq:SR_GGG} are defined as follows:
\begin{align}
&\mathcal{F}_1(n,a,b,l_{\text{min}},l_{\text{max}})
= \sum_{k=0}^{n} \frac{(-1)^k n!\,\Gamma(k+a)}{k!(n-k)!}
\nonumber\\
&\qquad
\times \sum_{l=l_{\text{min}}}^{l_{\text{max}}}
\frac{\Gamma(l+b)\Gamma(n-1-k+l)}{\Gamma(n-1+l+a)}
\nonumber \\
&\qquad \times \sum_{i=0}^{l} \frac{1}{i!(l-i)!(l-1-i+b)!}
\left(-\frac{m_c^2}{M^2}\right)^{l-i},
\\
&\mathcal{F}_2(n,a,b,l_{\text{min}},l_{\text{max}})
= \sum_{k=0}^{n} \frac{(-1)^k n!\,\Gamma(k+a)}{k!(n-k)!}
\nonumber\\
&\qquad
\times \sum_{l=l_{\text{min}}}^{l_{\text{max}}}
\frac{\Gamma(l+b)\Gamma(n-k+l)}{\Gamma(n+l+a)}
\nonumber\\
&\qquad
\times \sum_{i=0}^{l} \frac{1}{i!(l-i)!(l-1-i+b)!}
\left(-\frac{m_c^2}{M^2}\right)^{l-i},
\\
&\mathcal{H}(n,a,b) = \int_0^1 dx\, (2x-1)^n x^a \bar x^b
\exp\left[-\frac{m_c^2}{M^2\bar x}\right].
\end{align}
When applying the QCD sum rule expression for the decay constant $f_{D_s^+}$, the continuum threshold and the Borel parameter are crucial input parameters. In the conventional constraint scheme (scheme I), the contribution from the dimension-six condensate is typically suppressed to reduce the influence of higher-dimensional condensates on the calculated results. To further determine a reasonable range for the Borel parameter and improve the accuracy of the QCD sum rule calculation, we raise the criteria based on previous standards~\cite{Tian2023,Wang2025}, requiring that: a) the continuum contribution is less than $35\%$; b) the dimension-six condensate contribution does not exceed $1\%$; and c) the decay constant $f_{D_s^+}$ exhibits a stable dependence on $M^2$ within the Borel window. Accordingly, we select appropriate parameters according to the Borel parameter constraint criteria to obtain the $D_s^+$-meson decay constant.

In addition, we introduce a derivative constraint scheme (scheme II) to constrain the Borel window. The procedure is as follows. We first perform a simple transformation on Eq.~\eqref{Eq:SR_total} and define an auxiliary function $\xi$:
\begin{equation}
\xi\left(\frac{1}{M^2}\right) = f_{D_s^+}^2\left(\frac{1}{M^2}\right) \exp\left(-\frac{m_{D_s^+}^2}{M^2}\right).
\end{equation}
Setting $x = 1/M^2$, we have $\xi(x) = f_{D_s^+}^2(x) \exp(-m_{D_s^+}^2 x)$. We then construct the function $X_{m_{D_s^+}} = -\xi'/\xi$ (where $\xi' = d\xi/dx$, with the detailed expansion given in the Appendix), which simplifies to
\begin{equation}
X_{m_{D_s^+}} = m_{D_s^+}^2 - \frac{1}{f_{D_s^+}^2}\frac{d f_{D_s^+}^2}{dx}.
\end{equation}
Ideally, the decay constant $f_{D_s^+}$ should be independent of the Borel parameter, i.e., $df_{D_s^+}^2/dx \approx 0$, under which condition $X_{m_{D_s^+}} = m_{D_s^+}^2$. Finally, by choosing the continuum threshold according to the established criterion and imposing the condition $X_{m_{D_s^+}} = m_{D_s^+}^2$, we can determine the optimal Borel window. The key advantage of this scheme is that it does not require prior assumptions on the upper and lower bounds of the Borel window. Instead, it directly determines the range through the stability condition of the decay constant with respect to the Borel parameter. Moreover, the derivative operation further suppresses the contributions from higher-dimensional condensates, enhancing the accuracy of the extracted decay constant $f_{D_s^+}$.

\begin{figure}[t]
\centering
\includegraphics[width=0.35\textwidth]{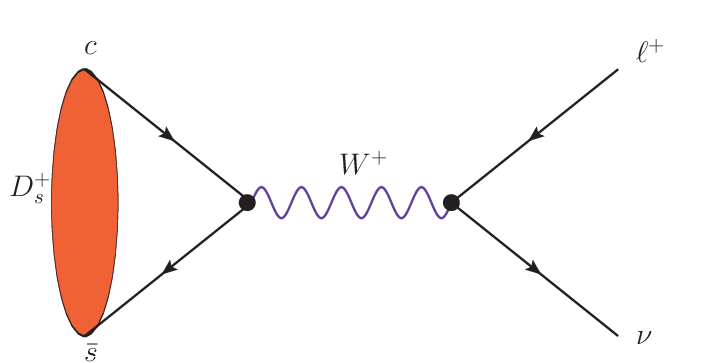}
\caption{Feynman diagram of the leptonic decay $D_s^+ \to \ell^+ \nu_\ell$ ($\ell = e, \mu, \tau$), where the $c$ and $\bar{s}$ quarks annihilate into a virtual $W^+$ boson.}
    \label{fig:feynman}
\end{figure}
Once the decay constant is determined, a theoretical prediction for the leptonic branching fraction can be made. As shown in Fig.~\ref{fig:feynman}, the $D_s^+$-meson annihilates into the $\ell^+\nu_\ell$ final state via the $c$ quark and $\bar{s}$ quark through a virtual $W^+$ boson. In the Standard Model, the branching fraction can be written as \cite{Marciano2004,Sirlin1982}
\begin{align}
\label{branchingratio}
\mathcal{B}(D_{s}^{+} \rightarrow \ell^{+}\nu_{\ell})
&= \frac{G_{F}^{2}}{8\pi} f_{D_{s}^{+}}^{2} m_{\ell}^{2} m_{D_{s}^{+}}
\bigg(1 - \frac{m_{\ell}^{2}}{m_{D_{s}^{+}}^{2}}\bigg)^{2}
\nonumber\\
&
\times |V_{cs}|^{2} \tau_{D_{s}^{+}} \left[1 + \frac{\alpha}{\pi} C_p\right],
\end{align}
where $G_F$ is the Fermi coupling constant, $m_\ell$ is the lepton mass, $m_{D_s^+}$ is the mass of the $D_s^+$-meson, $|V_{cs}|$ is the $D_s^+$-meson CKM matrix element and $\tau_{D_s^+}$ is the mean lifetime of the $D_s^+$-meson. The factor $(1 - m_\ell^2/m_{D_s^+}^2)^2$ originates from the two-body phase space integration. A larger charged-lepton mass $m_\ell$ leads to weaker helicity suppression. Consequently, the $e^+\nu_e$ channel is severely suppressed and difficult to measure experimentally. In practice, we typically use the branching fraction measurements of the $\mu^+\nu_\mu$ and $\tau^+\nu_\tau$ channels in our calculations. To obtain reliable predictions for the branching fractions, we also need to include the electroweak radiative corrections. The factor $[1 + (\alpha/\pi) C_p]$ accounts for the next-to-leading-order short-distance electroweak corrections, where the $C_p$ depends on the specific decay process. We apply the Sirlin short-distance electroweak correction of $-1.8\%$ to the branching fractions of both the $\mu^+\nu$ and $\tau^+\nu$ channels~\cite{Sirlin1982}. Finally, $|V_{cs}|$ can be extracted from the measured branching fraction using Eq.~\eqref{branchingratio} together with the theoretical decay constant.

{\it Numerical Analysis.--} In this section, we present the input parameters and numerical results for the QCD sum rule calculations. We adopt the charm quark mass $\bar{m}_c = 1.28 \pm 0.03\ \text{GeV}$, the strange quark mass $\bar{m}_s(2\ \text{GeV}) = 92.9^{+0.7}_{-0.7}\ \text{MeV}$, and the $D_s^+$ mean lifetime $\tau_{D_s^+} = (501.2 \pm 2.2) \times 10^{-15}\ \text{s}$, all taken from the PDG~\cite{PDG2026}. For the non-perturbative sector, the vacuum condensates are specified at the scale $\mu = 2\ \text{GeV}$ as follows~\cite{Zhong2021,Narison2015}:
\begin{align}
&\langle \bar{q}q \rangle = (-2.417^{+0.227}_{-0.114}) \times 10^{-2}\ \text{GeV}^3, \\
&\langle g_s\bar{q}\sigma T Gq \rangle = (-1.934^{+0.188}_{-0.103}) \times 10^{-2}\ \text{GeV}^5, \\
&\langle g_s\bar{q}q \rangle^2 = (2.082^{+0.734}_{-0.697}) \times 10^{-3}\ \text{GeV}^6, \\
&\langle \alpha_s G^2 \rangle = 0.037 \pm 0.011\ \text{GeV}^4, \\
&\langle g_s^3 f G^3 \rangle = 0.045\ \text{GeV}^6.
\end{align}
Since the above vacuum condensates and current quark masses are scale-dependent, they must be consistently evolved to the same scale via the renormalization group equations~\cite{Yang1993np,Hwang1994Delta} before being used in the calculations.
\begin{align}
&m_{s}(\mu) = m_{s}(\mu_{0}) \left[ \frac{\alpha_{s}(\mu)}{\alpha_{s}(\mu_{0})} \right]^{\frac{4}{9}}, \\
&\bar{m}_{c}(\mu) = \bar{m}_{c}(\bar{m}_{c}) \left[ \frac{\alpha_{s}(\mu)}{\alpha_{s}(\bar{m}_{c})} \right]^{-\frac{12}{25}}, \\
&\langle \bar{q}q \rangle (\mu) = \langle \bar{q}q \rangle (\mu_{0}) \left[ \frac{\alpha_{s}(\mu)}{\alpha_{s}(\mu_{0})} \right]^{-\frac{4}{9}}, \\
&\langle g_{s}\bar{q}\sigma T Gq \rangle (\mu) = \langle g_{s}\bar{q}\sigma T Gq \rangle (\mu_{0}) \left[ \frac{\alpha_{s}(\mu)}{\alpha_{s}(\mu_{0})} \right]^{\frac{2}{27}}, \\
&\langle g_{s}\bar{q}q \rangle^{2} (\mu) = \langle g_{s}\bar{q}q \rangle^{2} (\mu_{0}) \left[ \frac{\alpha_{s}(\mu)}{\alpha_{s}(\mu_{0})} \right]^{-\frac{4}{9}}, \\
&\langle \alpha_{s} G^{2} \rangle (\mu) = \langle \alpha_{s} G^{2} \rangle (\mu_{0}), \\
&\langle g_{s}^{3} f G^{3} \rangle (\mu) = \langle g_{s}^{3} f G^{3} \rangle (\mu_{0}).
\end{align}

\begin{figure}
  \centering
  \includegraphics[width=0.48\textwidth]{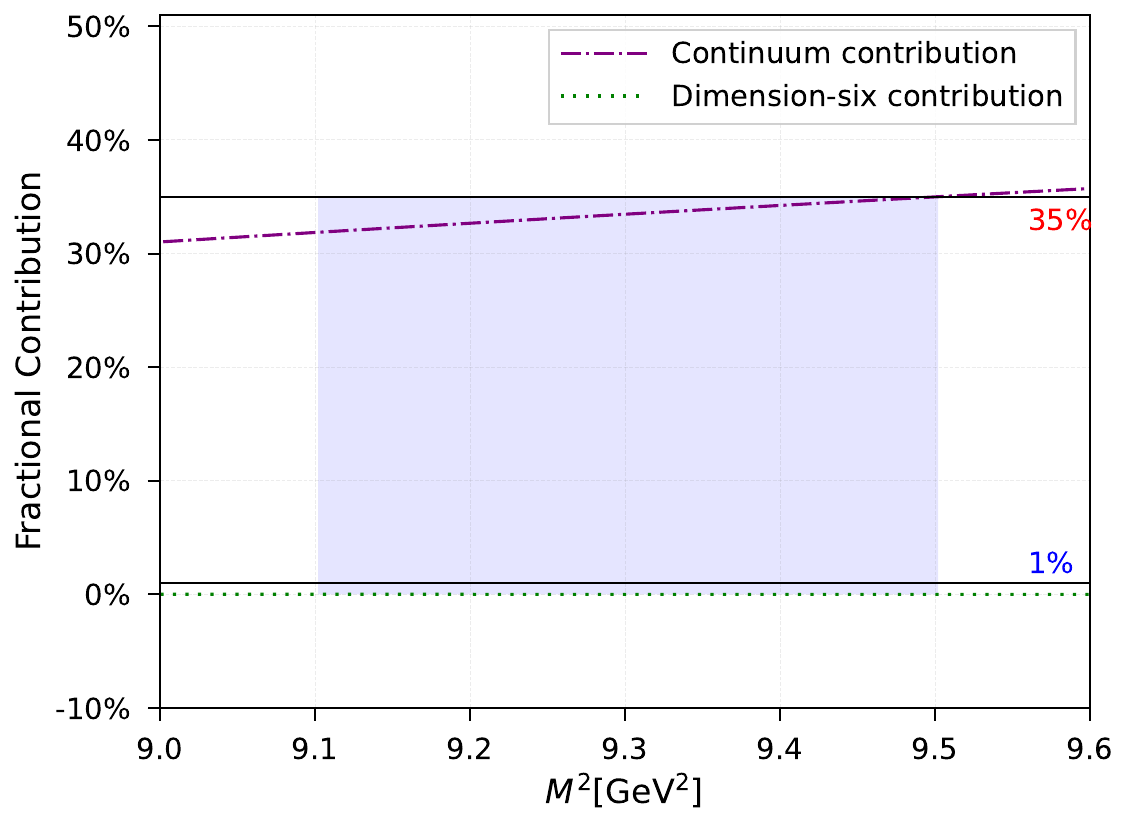}
  \caption{The continuum contribution and the dimension-six condensate contribution to the decay constant $f_{D_s^+}$ as functions of the Borel parameter $M^2$. The light shaded band indicates the Borel window $M^2$.}
  \label{fig:1}
\end{figure}
\begin{figure}
\centering
\includegraphics[width=0.48\textwidth]{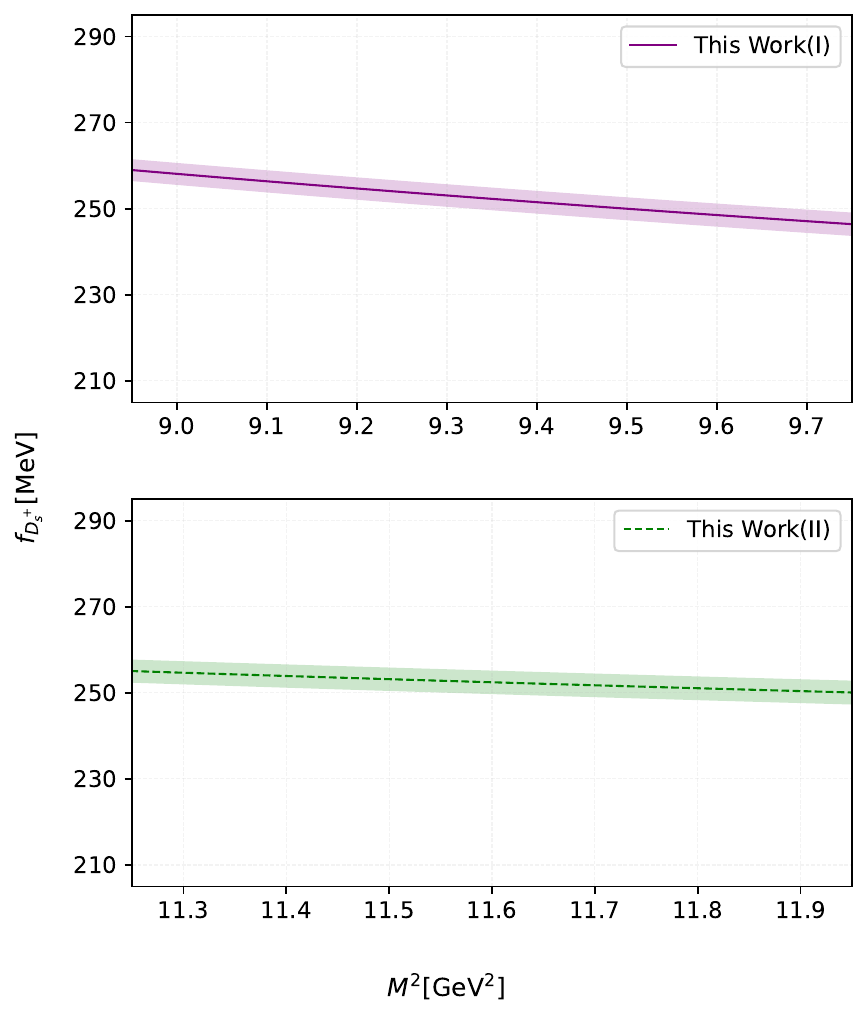}
\caption{The decay constant $f_{D_s^+}$ as a function of $M^2$ with its uncertainties in scheme I (upper) and scheme II (lower). The dashed lines represent the central values and the shaded bands indicate the uncertainties.}
\label{fig:2}
\end{figure}

Based on the above input parameters, we employ two different schemes to constrain the Borel window when calculating the decay constant $f_{D_s^+}$.
In the scheme I, we take the continuum threshold $s_0 = 6.7\ \text{GeV}^2$ near the squared mass of the first excited state $D_{s0}(2590)^+$ of the $D_s^+$-meson. Our numerical results indicate that the mass effect of the $s$-quark significantly suppresses the dimension-six condensate contribution, reducing its magnitude to the order of $10^{-5}$, which is far below the conventional criterion. This demonstrates that the contributions from higher-dimensional condensates are negligible for the precision of our calculation. Consequently, the lower bound of the Borel window cannot be directly determined by the conventional convergence criterion. To reduce the uncertainty, we take the value $0.2\ \text{GeV}^2$ below the upper bound as the reference central value of the Borel parameter. Fig.~\ref{fig:1} shows the continuum contribution and the dimension-six condensate contribution to the decay constant $f_{D_s^+}$ as functions of the Borel parameter $M^2$, where the light shaded band indicates the Borel window $M^2 \in [9.10,\ 9.50]\ \text{GeV}^2$. Within this window, the dependence of $f_{D_s^+}$ on $M^2$ is presented in Fig.~\ref{fig:2}, where the shaded band represents the uncertainties propagated from all input parameters. Thus, we obtain
\begin{equation}
f_{D_s^+}^{\text{(I)}} = 253.0^{+3.3}_{-3.1}\ \text{MeV}.
\end{equation}

In the scheme II, taking the derivative with respect to $x = 1/M^2$ modifies the weight of different terms in the OPE series, enhancing the perturbative contribution while significantly suppressing the dimension-six condensate contribution to the order of $10^{-6}$, which effectively reduces the uncertainty arising from higher-dimensional condensates. Following the criterion for the continuum threshold selection given in Ref.~\cite{Wang2015b}, we take $s_0 = 7.6\ \text{GeV}^2$. Combined with the above discussions, the Borel window is determined to be $M^2 \in [11.49,\ 11.89]\ \text{GeV}^2$ and the corresponding decay constant is obtained as
\begin{equation}
f_{D_s^+}^{\text{(II)}} = 251.8^{+1.4}_{-1.3}\ \text{MeV}.
\end{equation}
The detailed behavior of the results as a function of $M^2$, along with its uncertainty, is also shown in Fig.~\ref{fig:2}. Within the same Borel window width of $0.4\ \text{GeV}^2$, the variation of the decay constant is about $\pm 1.3\%$ for scheme I and about $\pm 0.6\%$ for scheme II. It is observed that after applying the scheme II, the decay constant exhibits significantly reduced dependence on the Borel parameter, indicating better stability.

\begin{figure}
\centering
\includegraphics[width=0.48\textwidth]{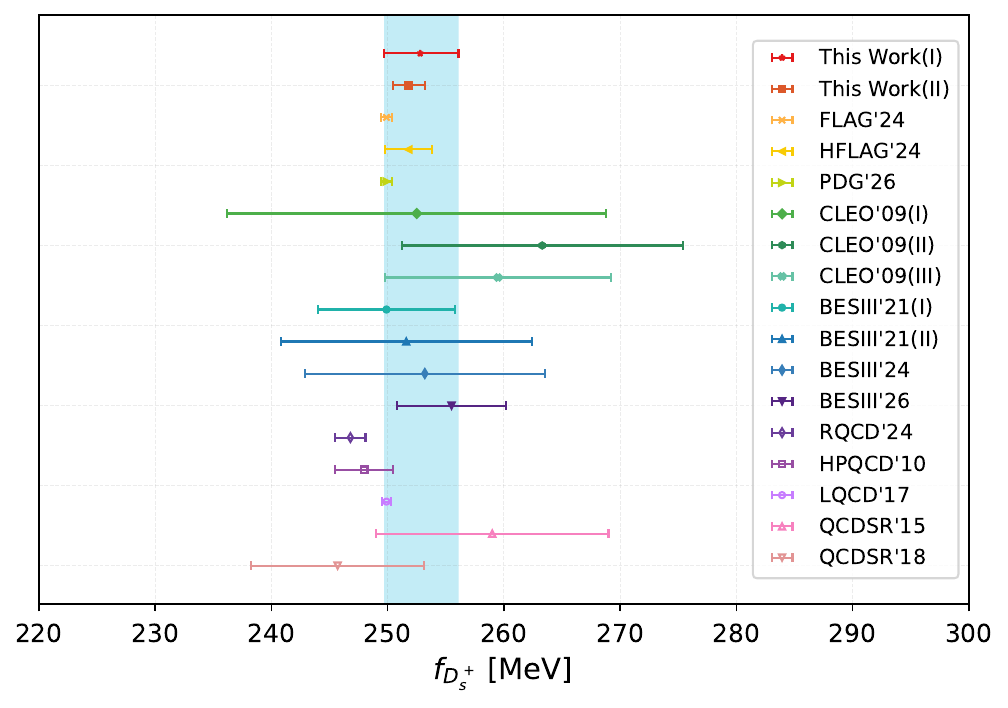}
\caption{Comparison of our results for $f_{D_s^+}$ with other theoretical and experimental values. The horizontal error bars indicate the uncertainties, which mainly arise from the input parameters.}
\label{fig:4}
\end{figure}

Furthermore, we present our results from two different schemes compared with experimental and theoretical predictions in Fig.~\ref{fig:4}. Especially for the FLAG'24~\cite{FLAG2411}, HFLAV'24~\cite{HFLAV2024}, PDG'26~\cite{PDG2026} for the average values, CLEO'09~\cite{CLEO0901a,CLEO0901b}, BESIII~\cite{BESIII2102,BESIII2105,BESIII2407,BESIII2606} for the experimental measurements, LQCD predictions from HPQCD'10~\cite{HPQCD2010}, Fermilab Lattice-MILC'17~\cite{LQCD1712}, RQCD'24~\cite{RQCD2405} are also included, as well as the QCD sum rule results from Wang'15~\cite{Wang2015a} and Mehraban'18~\cite{Mehraban2018Ds}. As can be seen from the figure, the experimental measurements have relatively large uncertainties dominated by statistical and systematic errors, while the lattice QCD predictions, ranging from about 246.8 to 249.9 MeV, are systematically lower. The central value of our scheme I result is in good agreement with CLEO'09(I)~\cite{CLEO0901a} and BESIII'24~\cite{BESIII2407}. In contrast, the scheme II result exhibits significantly reduced dependence on the Borel parameter and has a smaller central value, which is more compatible with the HFLAV'24~\cite{HFLAV2024} average and BESIII'21(II)~\cite{BESIII2105}. The observed spread among different experimental and theoretical groups suggests that a more precise determination of $f_{D_s^+}$ still requires a deeper understanding of non-perturbative effects, calling for further explorations of the $D_s^+$ decay constant.

\begin{table*}[t]
\footnotesize
\begin{center}
\caption{Comparison of our $D_s^+ \to \ell^+ \nu_\ell$ branching fractions with previous experimental and theoretical determinations.}
\label{tab:1}
\begin{tabular}{l l l l}
\hline
~~~~~~~~~~~~~~~~~~~~~~~~~~~~~~& $\mathcal{B}(D_s^+ \to e^+\nu_e)$~~~~~~~~~~~~ & $\mathcal{B}(D_s^+ \to \mu^+\nu_\mu) \times 10^{-3}$~~~~~~ & $\mathcal{B}(D_s^+ \to \tau^+\nu_\tau) \times 10^{-2}$ \\
\hline
This work(I)       & $(1.267_{-0.031}^{+0.033}) \times 10^{-7}$ & $5.390_{-0.130}^{+0.141}$ & $5.251_{-0.127}^{+0.137}$ \\
This work(II)      & $(1.255_{-0.013}^{+0.014}) \times 10^{-7}$ & $5.338_{-0.057}^{+0.061}$ & $5.200_{-0.056}^{+0.059}$ \\
HFLAV'24   ~\cite{HFLAV2024}    &---& $5.358 \pm 0.094 \pm 0.072$ & $5.362 \pm 0.064 \pm 0.073$ \\
PDG'26   ~\cite{PDG2026}    & $< 8.3 \times 10^{-5}$ & $5.37 \pm 0.11$ & $5.39 \pm 0.09$ \\
LQCD'19  ~\cite{Fleischer2019}& $(1.24 \pm 0.02) \times 10^{-7}$ & $5.28 \pm 0.08$ & $5.15 \pm 0.08$ \\
CLEO'09(I)   ~\cite{CLEO0901a}  &---& --- & $5.62 \pm 0.41 \pm 0.16$ \\
CLEO'09(II)  ~\cite{CLEO0901b}  & $< 1.2 \times 10^{-4}$ & $5.65 \pm 0.45 \pm 0.17$ &--- \\
BESIII'21(I)~\cite{BESIII2102} & --- & $5.35 \pm 0.13 \pm 0.16$ & $5.21 \pm 0.25 \pm 0.17$ \\
BESIII'21(II)~\cite{BESIII2105} & --- & --- & $5.29 \pm 0.25 \pm 0.20$ \\
BESIII'24~\cite{BESIII2407} & --- & $5.47 \pm 0.26 \pm 0.16$ & $5.60 \pm 0.16 \pm 0.20$ \\
BESIII'26~\cite{BESIII2606} & --- & --- & $5.37 \pm 0.08 \pm 0.06$ \\
\hline
\end{tabular}
\end{center}
\end{table*}

With the decay constants $f_{D_s^+}$ determined from the two constraint schemes, we proceed to compute the branching fractions for the leptonic decays $D_s^+ \to \ell^+ \nu$ ($\ell = e, \mu, \tau$), including the NLO electroweak radiative corrections. Our predictions are compared with other determinations in Table~\ref{tab:1}. For the $D_s^+ \to \mu^+\nu_\mu$ and $D_s^+ \to \tau^+\nu_\tau$ decay channels, which have larger lepton masses and thus larger branching fractions, they are more accessible experimentally. The available determinations for the $\tau$ channel range from about $5.15\%$ to $5.62\%$, and our predictions for this channel are in good agreement with the BESIII'21(I)~\cite{BESIII2102} within the uncertainties. The $\mu$ channel results are consistent with the latest PDG and HFLAV averages, so we adopt this channel to extract the CKM matrix element $|V_{cs}|$. For the $D_s^+ \to e^+\nu_e$ channel, our two determinations are consistent with the LQCD'19~\cite{Fleischer2019} prediction in central values. In practice, this channel is difficult to access due to helicity suppression, with only upper limits reported by experimental groups so far. Our predictions for this channel are expected to provide useful guidance for future experimental searches.

\begin{table*}[t]
\begin{center}
\footnotesize
\caption{Comparison of the $|V_{cs}|$ results obtained from different decay channels.}
\label{tab:2}
\begin{tabular}{l l l}
\hline
~~~~~~~~~~~~~~~~~~~~~~~~~~~~~~~~~~~~~~~~~~~~~& Decay channel ~~~~~~~~~~~~~~~~~~~~& $|V_{cs}|$ \\
\hline
This work(I) & $D_s^+ \to \mu^+ \nu_\mu$ & $0.967 \pm 0.012$ \\
This work(II) & $D_s^+ \to \mu^+ \nu_\mu$ & $0.970 \pm 0.005$ \\
FLAG'24  ~\cite{FLAG2411}   & --- & $0.974 \pm 0.002 \pm 0.011$\\
HFLAV'24  ~\cite{HFLAV2024}   & --- & $0.970 \pm 0.008$\\
PDG'26   ~\cite{PDG2026}    & --- & $0.969 \pm 0.005$ \\
BESIII'15 \cite{BESIII2015} & $D^+ \to \bar{K}^0 e^+ \nu_e$ & $0.975 \pm 0.008 \pm 0.015 \pm 0.025$ \\
BESIII'21(I)~\cite{BESIII2102} & $D_s^+ \to \mu^+ \nu_\mu$ & $0.973 \pm 0.012 \pm 0.015 \pm 0.004$ \\
BESIII'21(I)~\cite{BESIII2102} & $D_s^+ \to \tau^+ \nu_\tau$ & $0.972 \pm 0.023 \pm 0.016 \pm 0.004$ \\
BESIII'21(II)~\cite{BESIII2105} & $D_s^+ \to \tau^+ \nu_\tau$ & $0.980 \pm 0.023 \pm 0.019$ \\
BESIII'24~\cite{BESIII2407} & $D_s^+ \to \mu^+ \nu_\mu$ & $0.986 \pm 0.023 \pm 0.014 \pm 0.003$ \\
BESIII'26~\cite{BESIII2606} & $D_s^+ \to \tau^+ \nu_\tau$ & $0.993 \pm 0.008 \pm 0.006 \pm 0.003 \pm 0.003$ \\
HPQCD'10 ~\cite{HPQCD2010}  & --- & $1.010 \pm 0.022$\\
RBC/UKQCD'17 ~\cite{Boyle2017}  & --- & $1.011 \pm 0.016$ \\
\hline
\end{tabular}
\end{center}
\end{table*}

Subsequently, we use the latest branching fraction $\mathcal{B}(D_s^+ \to \mu^+ \nu_\mu)$ from the PDG~\cite{PDG2026} as input. The uncertainty in the extracted $|V_{cs}|$ is mainly determined by the input parameters. In Table~\ref{tab:2}, we compare our results with those from various experimental and theoretical groups. The early lattice QCD predictions are around $1.010$, and the BESIII measurements from 2024 and 2026 also show an increasing trend compared with the world averages. In contrast, the latest averages from FLAG, HFLAV and PDG are much closer to each other. The extracted values obtained from our two schemes agree with the PDG'26~\cite{PDG2026} and HFLAV'24~\cite{HFLAV2024} within errors.

{\it Summary.--}Based on the background field theory, we have calculated the $D_s^+$-meson decay constant $f_{D_s^+}$ using the QCD sum rules approach, with the complete expression derived up to dimension-six condensates. To determine the Borel window, we adopt two different constraint schemes. In scheme I, following the conventional Borel window criteria, we obtain $f_{D_s^+}^{\text{(I)}}$. In scheme II, we apply the derivative with respect to $x = 1/M^2$ under mass constraints, effectively suppressing the dependence of the decay constant on the Borel window and yielding $f_{D_s^+}^{\text{(II)}}$. Using these results, we predict the branching fractions of the $D_s^+ \to \ell^+ \nu_\ell$ decays, including the NLO electroweak radiative corrections. The predictions for the $\mu$ and $\tau$ channels are in good agreement with the experimental measurements. For the $e$ channel, which is difficult to measure experimentally, our predictions are consistent with the current LQCD results~\cite{Fleischer2019}. Besides, combined with the latest branching fraction from the PDG, we extract the CKM matrix elements $|V_{cs}|^{\text{(I)}}$ and $|V_{cs}|^{\text{(II)}}$ from the two schemes, respectively. Both values are consistent with the latest PDG and HFLAV averages~\cite{PDG2026,HFLAV2024} within errors. We hope our predictions could provide useful references for further studies of purely leptonic $D_s^+$ decays.\\

\textit{Acknowledgments.--}The research was supported by the National Natural Science Foundation of China under Grant No.12265010, the Project of Guizhou Provincial Department of Science and Technology under Grants No. MS[2025]219 and No. CXTD[2025]030.\\

\appendix
\textit{Appendix.--} In the QCD sum rule expression obtained by the derivative method in this work, the perturbative contribution and the individual nonperturbative condensate contributions are expressed as follows:
\begin{widetext}
\begin{align}
&\xi'(s,M^2) = I'_{\text{pert}}(s,M^2) + I'_{\langle \bar{q}q \rangle}(M^2) + I'_{\langle G^2 \rangle}(M^2) + I'_{\langle \bar{q}Gq \rangle}(M^2) + I'_{\langle \bar{q}q \rangle^2}(M^2) + I'_{\langle G^3 \rangle}(M^2),
\label{Eq:fDsprime_tot}
\\
&I'_{\rm pert}(s,M^2) = \int_{l_{\text{min}}}^{s_0^{D_s^+}} -\frac{ds}{16\pi^2 s^2} \,e^{-\frac{s}{M^2}}
\{(m_s^2 - m_c^2 + {\cal C}{\cal D}) [ 3s^2 - (m_s^2 - m_c^2 - {\cal C}{\cal D})^2]-({\cal C}{\cal D} \leftrightarrow  - {\cal C}{\cal D})\},
\label{Eq:fDsprime_pert}
\\
&I'_{\langle \bar{q}q \rangle}(M^2) = e^{-\frac{m_c^2}{M^2}} \frac{(M^2 - m_c^2) m_q \langle \bar{q}q \rangle}{M^2},
\label{Eq:fDsprime_qq}
\\
&I'_{\langle G^2 \rangle}(M^2) =
\frac{\langle \alpha_s G^2 \rangle}{12 \pi}
\int_0^1 dx\, e^{-\frac{m_c^2}{M^2 \bar{x}}}
\Bigg(
1- \frac{m_c^2}{M^2} \frac{x+1}{\bar{x}^2}+ \frac{m_c^4}{M^4} \frac{x}{\bar{x}^3}
\Bigg),
\label{Eq:fDsprime_GG}
\\
&I'_{\langle \bar{q}Gq \rangle}(M^2) = e^{-\frac{m_c^2}{M^2}}
\frac{m_c^2 m_q \langle g_s \bar{q}\sigma\cdot Gq \rangle}{18 M^6} (4m_c^2 - 3M^2),
\label{Eq:fDsprime_qGq}
\\
&I'_{\langle \bar{q}q \rangle^2}(M^2) = e^{-\frac{m_c^2}{M^2}}
\frac{2\langle g_s \bar{q}q \rangle^2}{81 M^4}(2M^2-m_c^2),
\label{Eq:fDsprime_qq2}
\\
&I'_{\langle G^3 \rangle}(M^2) =
e^{-\frac{m_c^2}{M^2}} \frac{\langle g_s^3 fG^3 \rangle}{288 M^4 \pi^2}
\Bigg\{51 m_c^2 \left( \frac{11}{6}  \, - \,  \gamma_E + \ln\frac{M^2}{\mu^2} \right)  \, +  \, 2 \sum_{l=1}^{50} \Gamma(l)
 \sum_{i=0}^{l} \frac{(i-l) m_c^2 \left(-\dfrac{m_c^2}{M^2}\right)^{l-1-i}}{i! (l-i)! (l-i+2)!} \, - \,  3 \sum_{l=1}^{50}
\notag
\\
&\qquad \qquad~ \times \frac{\Gamma(l)\Gamma(l+3)}{\Gamma(l+4)} \sum_{i=0}^{l} \frac{(i-l) m_c^2 \left(-\dfrac{m_c^2}{M^2}\right)^{l-1-i}}{i! (l-i)! (l-i+2)!}
- 51 \sum_{l=2}^{50} \frac{\Gamma(l-1)\Gamma(l+3)}{\Gamma(l+4)}
\sum_{i=0}^{l} \frac{(i-l) m_c^2 \left(-\dfrac{m_c^2}{M^2}\right)^{l-1-i}}{i! (l-i)! (l-i+2)!}
\notag
\\
&\qquad \qquad~ - m_c^2 \Bigg[ \frac{17 m_c^2}{M^2} + 2 \sum_{l=1}^{50} \Gamma(l) \sum_{i=0}^{l} \frac{\left(-\dfrac{m_c^2}{M^2}\right)^{l-i}}{i! (l-i)! (l-i+2)!} - 3 \sum_{l=1}^{50} \frac{\Gamma(l)\Gamma(l+3)}{\Gamma(l+4)}
\sum_{i=0}^{l} \frac{\left(-\dfrac{m_c^2}{M^2}\right)^{l-i}}{i! (l-i)! (l-i+2)!} - 51
\notag
\\
&\qquad \qquad~  \times \sum_{l=2}^{50} \frac{\Gamma(l \! - \! 1)\Gamma(l \! + \! 3)}{\Gamma(l \! + \! 4)}
\sum_{i=0}^{l} \frac{\left(-\dfrac{m_c^2}{M^2}\right)^{l-i}}{i! (l-i)! (l-i+2)!} + \left( 50  \! + \!  \frac{51 m_c^2}{M^2} \right) \left( \frac{3}{2}  \! -  \! \gamma_E  \! + \!  \ln\frac{M^2}{\mu^2} \right)
\Bigg] \Bigg\} + \frac{\langle g_s^3 fG^3 \rangle m_c^2}{1440 M^8 \pi^2}
\notag
\\
&\qquad \qquad~ \times \Bigg[ \int_0^1 dx \, \frac{e^{-\frac{m_c^2}{M^2 \bar{x}}}}{\bar{x}^2} \,
\left( \, \frac{10 M^8 x}{2x-1}  \, + \,  5 M^4  \, - \,  20 m_c^2 \right)   \, +  \,  \int_0^1 dx \frac{e^{-\frac{m_c^2}{M^2 \bar{x}}}}{\bar{x}^3}
\left(\frac{10 M^4 m_c^2 x}{2x-1} + 10 m_c^2 + 240 M^4 x
\right)
\notag
\\
&\qquad \qquad~  + 15 m_c^2 \int_0^1 dx\, \frac{e^{-\frac{m_c^2}{M^2 \bar{x}}} x}{\bar{x}^4} - 6 \frac{m_c^4}{M^4} \int_0^1 dx\, \frac{e^{-\frac{m_c^2}{M^2 \bar{x}}} x}{\bar{x}^5}
\Bigg]
\label{Eq:fDsprime_G31}
\end{align}
\end{widetext}

\end{document}